\documentclass[aps,prl,superscriptaddress,twocolumn,showpacs,floatfix]{revtex4}
\usepackage{graphicx}
\usepackage{dcolumn} 
\usepackage{color}
\usepackage{wrapfig}

\begin{document}
\bibliographystyle{apsrevnourl}
\title{Reliability of genetic networks is evolvable} 
\author{Stefan Braunewell}
\author{Stefan Bornholdt} 
\affiliation{Institute for Theoretical Physics, University of Bremen, 
D-28359 Bremen, Germany} 
\date{\today}
\begin{abstract}
Control of the living cell functions with remarkable reliability despite the stochastic 
nature of the underlying molecular networks --- a property presumably optimized 
by biological evolution. We here ask to what extent the property of
a stochastic dynamical network to produce reliable dynamics is an evolvable trait. 
Using an evolutionary algorithm based on a deterministic selection criterion
for the reliability of dynamical attractors, we evolve dynamical networks of
noisy discrete threshold nodes. We find that, starting from any random network, 
reliability of the attractor landscape can often be achieved with only few small 
changes to the network structure. Further, the evolvability of networks towards 
reliable dynamics while retaining their function is investigated and a high 
success rate is found.
\end{abstract}
\maketitle
The processes of life in cells and organisms are largely controlled by 
complex networks of molecular interactions as, for example, networks 
of regulatory genes. A remarkable feature of these networks is their 
reliable functioning, despite their molecular components being 
subject to noise of both, intrinsic as well as extrinsic nature 
\cite{Rao:2002fk,TO2001}. How does the interplay of such unreliable 
components ensure a reliable functioning of the networks that control 
cells and organisms? 

Naturally, properties of the circuitry can be expected to play a major role, 
and indeed some topological features of regulatory networks, such as
feedback loops and gene redundancy, are known to aid robustness 
of noisy systems \cite{McAdams:1999}. Further studies found numerous 
evidence for a close interplay of topology and robustness of networks 
\cite{SMMA2002,Aldana:2003,KB2004}. What is the origin of 
such reliable network structures? 

Starting from the fact that real-world biological systems are the result of 
evolutionary processes, noise resistance of biological networks 
presumably emerged from the interplay of mutation and selection, as well. 
We here study the question of how accessible noise resistant dynamical 
networks are to evolution and what the costs in terms of topological 
rearrangements are in order to achieve a reliable dynamical network. 
We study this question in the framework of numerical experiments, 
evolving discrete dynamical networks in the computer. 

Evolving genetic networks in the computer has a long tradition 
\cite{KauffmanSmith86,Wagner:1997,BS1998}. Several concepts 
of robustness have been studied in this framework, reaching from 
robustness of network dynamics against mutational perturbations 
\cite{Wagner:1997}, to robustness of expression patterns during 
evolution (neutral evolution) \cite{BS1998}, 
as well as robustness of attractors against switching errors 
of genes \cite{szejka-2007,Ciliberti:2007lr}.  

In this paper we extend these viewpoints by studying the evolution 
of networks towards robustness against small timing fluctuations or
``reliability'' (to avoid confusion with existing definitions of robustness). 
While gene switching errors (a type of ``perturbation'' very commonly 
used by many authors) are not exactly small perturbations, 
and may not be the common case in a real cell, small perturbations 
in timing and activity levels are ubiquitous in biological systems. 
Such small noise levels have recently proven to destroy most 
attractors in Boolean networks that are observed under parallel 
update \cite{greil:048701,KB2004-2}. Obviously, only those 
attractors that are stable against such small noise (i.e., ``reliable'')
can be relevant in the biological context. Indeed, in the biological 
example of the yeast cell cycle network, this type of stability against
timing perturbations is observed \cite{Braunewell:2007lr}.

Here, we investigate whether such reliability of a dynamical network 
can readily result from an evolutionary procedure. Defining biologically 
motivated mutation-selection processes, we will evolve random networks 
towards realizations that exhibit reliable dynamics. 
We investigate both the emergence of fully stable attractor landscapes
as well as the ability of networks to evolve in such a way that a given 
attractor is stabilized.

We model genes as nodes in a network, where the links between two nodes determine the interactions between the genes. All bio-molecular processes are simply substituted by such a link. The presence of a gene's transcript is modeled as a simple on-off switch, the state of which is called ``activity state''. A node can have several inputs and in principle the activity state of a node can depend on its inputs through any Boolean rule. As we will use an evolutionary process to find robust networks, we wish to simplify the rules such that the dynamics is fully determined by the network structure alone, with no additional freedom of choice in the rules. Thus, we choose as a suitable subset of possible Boolean networks a threshold network, which amounts to a majority rule in the inputs of each node. Every node has a state of either $1$ (active) or $-1$ (inactive). We allow the links to carry a weight of either $+1$ or $-1$, corresponding to an activating or inhibiting interaction, respectively. The update rule in the synchronous case is given by:
\begin{equation}
S_i(t+1)= \left \{
\begin{array}{l}
	+1 \qquad \mathrm{if} \, \sum_{i=1}^n A_{ij} S_j(t) \geq 0, \\
	-1 \qquad \mathrm{otherwise},
\end{array}
\right.
\label{eupdaterule}
\end{equation}
where $A_{ij}$ characterizes the link from node $j$ to node $i$ and $S_i$ denotes the state of node $i$.

As the dynamics is discrete (finite state space) and deterministic, for every initial condition the system reaches an attractor, which can be either a fixed point ($S_i (t+1) = S_i (t) \, \forall i$) or a limit cycle, where the same sequence of states is repeated indefinitely. 

To assess the stability of a network against fluctuations of the signal times, we use the stability criterion of \cite{KB2004-2} which provides a deterministic measure for a network under investigation. 
It requires two principle assumptions: the nodes implement a low-pass filter that removes the effect of activity states that are maintained only over short time spans; and the signal time fluctuations are small compared to the time scales of the processes and that of the filter.

The first assumption is justified by the buildup and decay processes of protein concentrations \cite{Hi2002}. Gene activity states that persist only for a short time do not significantly affect other proteins as the gene's transcript can only be produced in small numbers.
The second assumption means that we are investigating systems with low noise. A single signal fluctuation does not significantly perturb the system, but only the addition
of many similar perturbations over time can drive the system away from 
an initially synchronous behavior.

We now give the formal description of the systematic stability test:
To determine the stability of an attractor, first the synchronous state sequence is determined by full enumeration of initial states. Choosing one step in an attractor, we determine all switches that occur at this step and call the set of switching nodes $M$. For every proper, non-empty subset $S\subset M$ we change the switching times from $t=0$ to $t=\epsilon$, i.e. we retard the switching times for these nodes by an infinitesimal number. Thus, a new intermediate state from time $t=0$ to $t=\epsilon$ is created, where some nodes have already switched, whereas other nodes still exhibit the state of the previous (synchronous) time step. We then follow the dynamics, with two times for every synchronous time step:

1. Determine the states at times $t=i$, $i=1,2,\dots$ and $t'=i+\epsilon$ from the states at $t=i-1$ and $t'=i-1+\epsilon$, respectively.

2. Apply the filter rule: if a node switches both at integer and perturbed time, remove both switches. As the activity state has persisted only for a time span of $\epsilon$ we assume it does not further affect the system.

3. If all nodes switch at either integer or perturbed time, the system has regained synchrony and the attractor is stable against this particular subset of perturbed nodes. If however, the system reaches a new attractor in the combined state space of both times, the system is unstable as the perturbation can in general persist in the system and might diverge, thus leading to a different atttractor or to ``chaotic'' regime of incessant switchings.

We call an attractor ``stable'' if it is stable against all subset perturbations, otherwise we call it ``unstable''. Fixed points are trivially stable by this definition.

We use an evolutionary algorithm to drive the networks to stability. In every step, the network is mutated and the result of the stability assessment is compared with the mother network. If the mutant fitness is higher than that of the original network, the mutant is kept and replaces the original, otherwise, a new mutant is tested. This is repeated until the requested criterion is fulfilled. As different selection criteria are used, the definition of the fitness score is given in the respective part of the results section.

Mutation is performed through a single link rewiring, which means that at the same time a connection between two nodes is removed and a new connection between two nodes is added. This procedure amounts to two elementary manipulations of the network structure but it has the advantage that the average connectivity of the network is unchanged by the mutation. This allows for better comparison between the random and the evolved networks. As our method requires full enumeration of the space of $2^N$ states where $N$ is the number of nodes, we can only perform this analysis for small networks. We show the results for $N=16$ nodes, but have checked that the conclusions also hold for networks with $N=12$ and $N=20$ nodes.

In the first part let us evolve networks towards stability regarding the complete attractor landscape. We define the evolution process in the following way: given a network, we accept a mutation of it, if the mutant has a higher number of initial states leading to a stable attractor. If so, the network is replaced by the mutant and the next evolution step is taken, otherwise a new mutation is tested. This procedure stops as soon as all initial states lead to stable attractors.

\begin{figure}
\includegraphics[angle=-90,width=8cm]{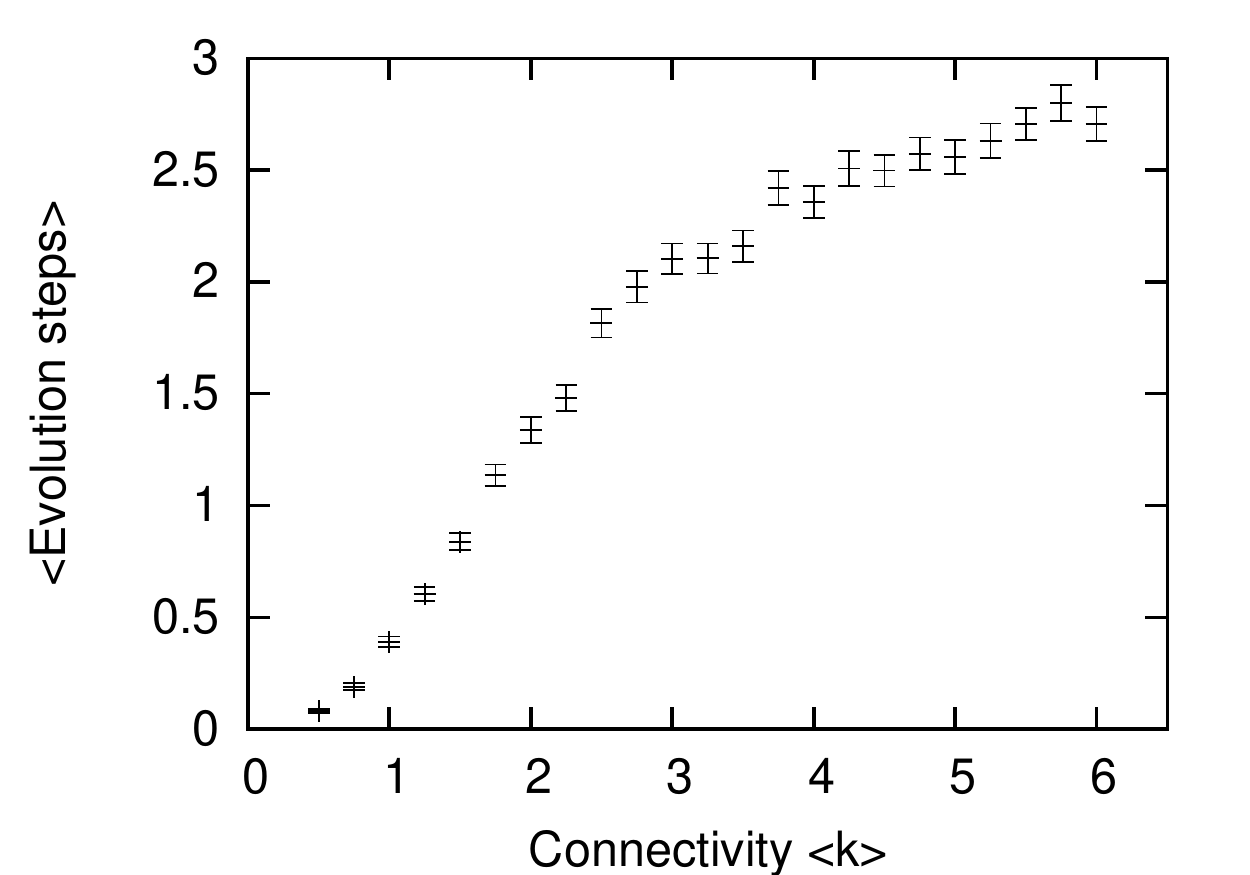}
\caption{Network evolution rapidly leads to stable attractor landscapes: average number of evolution steps vs. average connectivity of the networks. The networks consist of 16 nodes, each data point corresponds to the average of 1000 runs starting with random networks of the respective connectivity.}
\label{fevovsconn}
\end{figure}

In figure \ref{fevovsconn} we show the average number of evolution steps necessary to reach full stability of the attractor landscape, plotted against the average connectivity, defined by the total number of edges divided by the number of nodes. Networks consist of 16 nodes and 1000 repetitions were run for every data point. One can see that for all connectivities a very small number of mutations already suffices to find a completely stable network. Using a more restrictive method of selection, like choosing the fittest out of several tested mutant networks, further reduces the average evolution steps significantly (data not shown).

\begin{table}
\begin{center}
\begin{tabular}{|l|r|r|}
\cline{1-3}
& random networks & evolved networks \\
\cline{1-3}
number of attractors & $3.98 \pm 0.02$ & $2.12 \pm 0.01$\\
largest basin size & $47800 \pm 100$  & $57100 \pm 100$ \\
IS to stable atts. & $40300 \pm 200$ & 65536 \\
\# of evolution steps & -- & $2.07 \pm 0.02$ \\
\cline{1-3}
\end{tabular}
\end{center}
\caption{Comparison of attractor basin characteristics of random and evolved networks for N=16, $\langle k\rangle=3$. Averages over 20000 runs.}
\label{toriginal}
\end{table}

Next, it is interesting to look at network properties and how they change during the course of the evolution process. In table \ref{toriginal} we compare random networks with networks that have undergone the evolution process for an average connectivity of $\langle k\rangle=3$ (but the qualitative results are typical for any value of $\langle k\rangle$). One can see that the average number of attractors has decreased and that the size of the largest basin has increased at the same time. 
Again, these significant effects take place within very few evolution steps. Thus, we find that the dynamical landscape of a threshold network can be significantly altered by only a few mutations of the network topology. Stability of the attractor landscape can be achieved without significant changes of the overall network structure. 

To ensure that we do not simply observe the effects of networks evolving towards fixed points (which are always stable), we have checked all results also with the rule that a fixed point is counted as an unstable attractor. We do not show the results here, but the general results and the conclusions drawn above hold also in this case.

These results clearly show that in our simple dynamical model, random networks exhibit an astonishing evolvability regarding their attractor landscapes. The landscape can be easily shaped and robustness in dynamical functioning can be achieved after very few evolution steps. 

So far we have demonstrated that random networks can be quickly evolved towards a completely stable attractor landscape. However, we have not constrained the dynamics in any way, so the evolved networks might show completely different dynamical behavior than the original networks. If we think of attractors as a function being performed by a genetic network, we should restrict evolution to those networks that are able to reproduce the original attractor dynamics. 

This leads to a modified selection criterion with the following target: We choose the largest attractor of the original network as the ``functional attractor'' and require stabilization of this attractor. If it is a fixed point or a stable limit cycle, there is trivially nothing to do in the evolution, so we just discard these networks and create a new one until we find a network with an unstable largest attractor. During evolution, every mutant has to reproduce this attractor. This means that, starting at one step of the attractor cycle, the dynamics of the original network and of the mutant have to be exactly the same. If the mutant does not reproduce the attractor, it is immediately discarded. We do not request the networks to reproduce the transient states as this constraint is too strict and disallows practically every mutation.

The fitness score is given by the multiplication of the stability value (0 if unstable, 1 if stable) with the basin size of the functional attractor. We have employed two different selection criteria: strict or neutral selection. In the strict selection scheme, a network is only accepted, if it increases the fitness score, whereas in the neutral selection a larger or equal fitness score suffices. This means that in the strict
scheme, the stabilization has to occur within a single rewiring, whereas the neutral criterion allows for 
a random walk through the space of networks that exhibit the functional attractor.
The evolution process is complete as soon as the functional attractor is stable with a basin size of half the total state space, which makes the functional attractor the dominant dynamical expression pattern. 

\begin{figure}
\includegraphics[angle=-90,width=8cm]{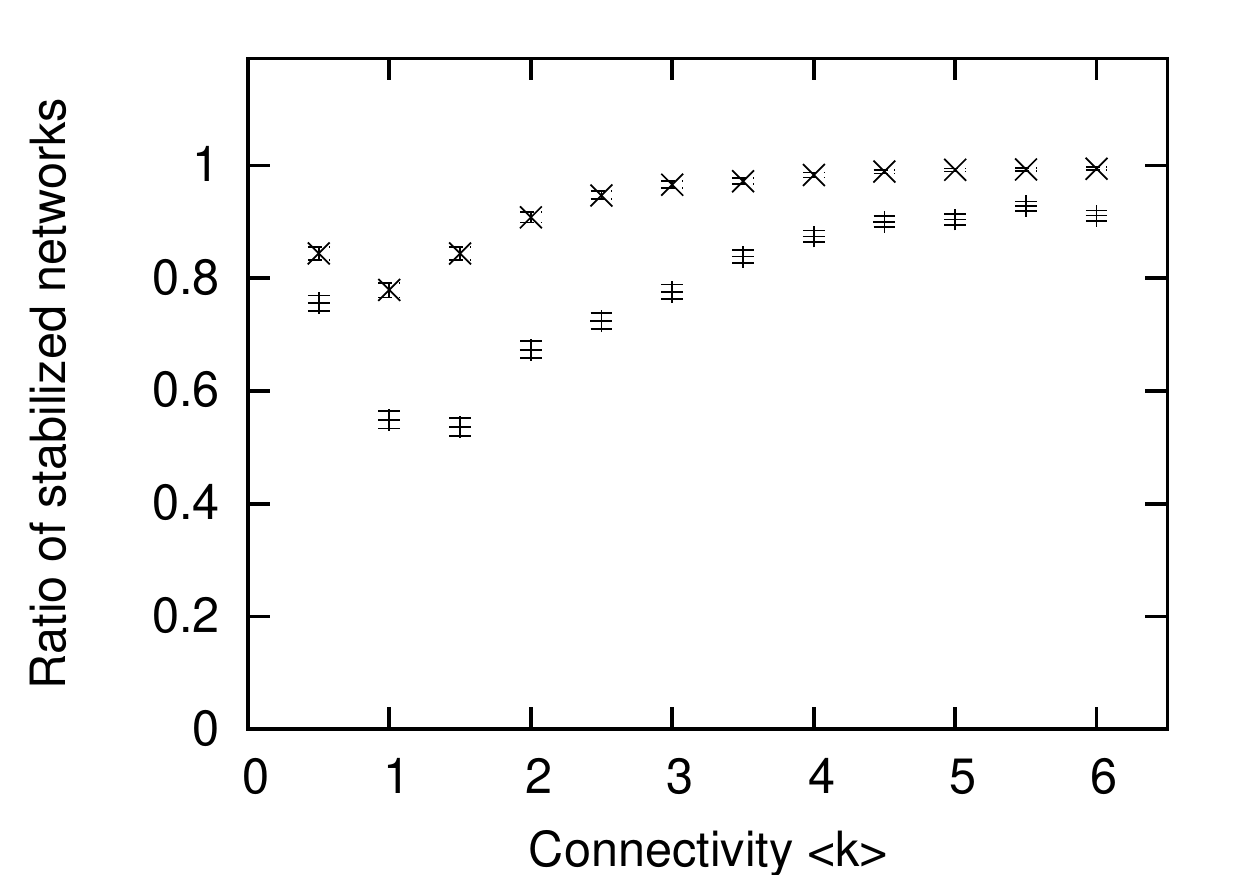}
\caption{Functional attractor stabilization. The ratio of networks successfully stabilized in the evolution is plotted against average connectivity. Network size $N=16$, every data point amounts to the average of 1000 runs at the respective connectivity. Upper points ($\times$): neutral mutations (see text). Lower points (+): single link rewirings. Many networks can be stabilized, with neutral mutations practically all networks with connectivities larger than 2.}
\label{fratiovsconn}
\end{figure}

In figure \ref{fratiovsconn} we show the results of the evolution processes using the functional attractor criterion for a network size of $N=16$ and 1000 attempted evolution runs. The ratio of networks that can be stabilized in both selection schemes is plotted against the average connectivity of the networks. For each evolution step, we have attempted $20000$ mutations before marking a network as not evolvable towards stability (this arbitrary criterion does not influence the results as long as the number of attempts is sufficiently high). In the neutral selection, a stable network has to be found within $10^6$ mutation attempts during the full evolution run.

First, one can see that even in the single-step evolution (marked by a cross $\times$), more than half of all networks can be stabilized. For very low connectivity as well as connectivities above 3, more than 3/4 of all networks fulfill the criterion. The dip at connectivities around $1.5$ can be attributed to the fact that network dynamics start to become complex, but the possibility of affecting the stability without destroying the attractor are small. At higher connectivities, the degeneracy of the topology aids the stabilization of the system, at small connectivities, self-couplings can often create a stable dynamical core which drives
all other nodes like a central clock.

Next, we investigate the effect of the neutral selection.  
The chance of finding a network with stabilized functional attractor is significantly increased
as compared to the single-step stabilization. Especially for networks of connectivities around $1.5$, the additional number of stabilized networks is high. For connectivities above 2, practically every network can be stabilized using this evolution process.

\begin{figure} 
\includegraphics[width=8.5cm]{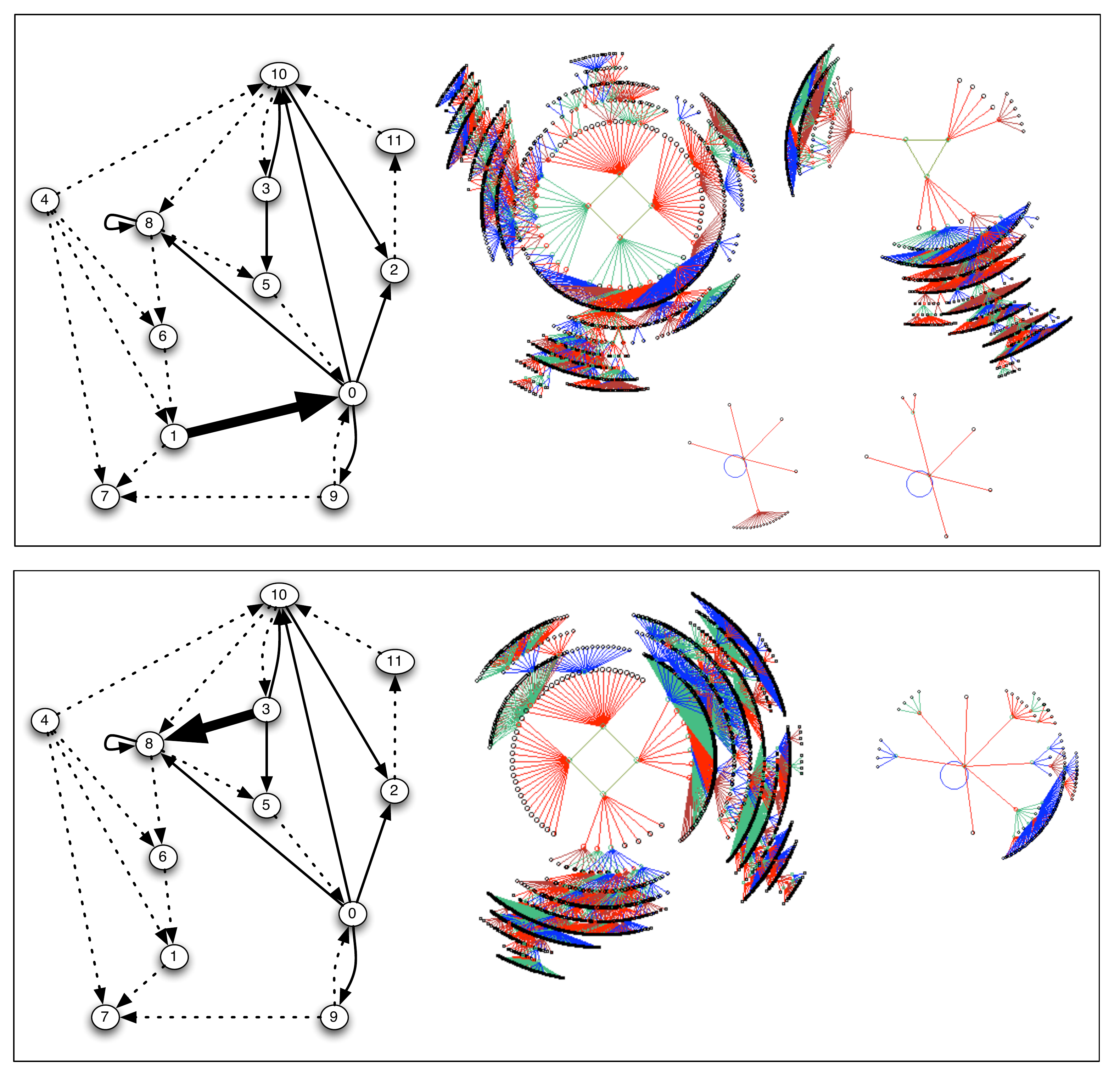} 
\caption{A single rewiring can dramatically affect the attractor landscape. The network structure and attractor landscape (state space visualization) of a network before (top) and after (bottom) a single rewiring. Central shape in the attractor pictures show limit cycle, transient states are arranged on arcs.}
\label{fattlandscape} 
\end{figure} 

In figure \ref{fattlandscape} we show a typical example of an evolutionary process for a network with 12 nodes and $\langle k\rangle=2$ (created with DDLAB \cite{Wuensche:1994}). The network structure as well as the full synchronous attractor landscape is shown, before (top) and after a single mutation (bottom). In the attractor landscape figure, each network state is represented by a dot that is connected to the concurrent state by a line. The central shape denotes the limit cycle (or fixed point). All four attractors of the original network are unstable. After mutation (mutated links are depicted by thick lines), only two attractors remain. The functional attractor with a cycle length of four is now stable. One can see how the single mutation dramatically affects the attractor landscape of the network. 

Two implications of our results are at hand. First, we find that the topological features of a network do not strictly constrain the stability of the resulting network dynamics. Small changes in the rewiring can have dramatic effects on the attractor landscape, including complete stabilization. Second, the (synchronous) state sequence of an attractor does not determine the stability. Even within small topological changes, it is often possible to find networks which exhibit the same attractor, but perform it in a reliable way.

\clearpage


\begin{thebibliography}{16}
\expandafter\ifx\csname natexlab\endcsname\relax\def\natexlab#1{#1}\fi
\expandafter\ifx\csname bibnamefont\endcsname\relax
  \def\bibnamefont#1{#1}\fi
\expandafter\ifx\csname bibfnamefont\endcsname\relax
  \def\bibfnamefont#1{#1}\fi
\expandafter\ifx\csname citenamefont\endcsname\relax
  \def\citenamefont#1{#1}\fi
\expandafter\ifx\csname url\endcsname\relax
  \def\url#1{\texttt{#1}}\fi
\expandafter\ifx\csname urlprefix\endcsname\relax\def\urlprefix{URL }\fi
\providecommand{\bibinfo}[2]{#2}
\providecommand{\eprint}[2][]{\url{#2}}

\bibitem[{\citenamefont{Rao et~al.}(2002)\citenamefont{Rao, Wolf, and
  Arkin}}]{Rao:2002fk}
\bibinfo{author}{\bibfnamefont{C.~V.} \bibnamefont{Rao}},
  \bibinfo{author}{\bibfnamefont{D.~M.} \bibnamefont{Wolf}}, \bibnamefont{and}
  \bibinfo{author}{\bibfnamefont{A.~P.} \bibnamefont{Arkin}},
  \bibinfo{journal}{Nature} \textbf{\bibinfo{volume}{420}},
  \bibinfo{pages}{231} (\bibinfo{year}{2002}).

\bibitem[{\citenamefont{Thattai and van Oudenaarden}(2001)}]{TO2001}
\bibinfo{author}{\bibfnamefont{M.}~\bibnamefont{Thattai}} \bibnamefont{and}
  \bibinfo{author}{\bibfnamefont{A.}~\bibnamefont{van Oudenaarden}},
  \bibinfo{journal}{Proc.~Natl.~Acad.~Sci.~USA} \textbf{\bibinfo{volume}{98}},
  \bibinfo{pages}{8614} (\bibinfo{year}{2001}).

\bibitem[{\citenamefont{McAdams and Arkin}(1999)}]{McAdams:1999}
\bibinfo{author}{\bibfnamefont{H.~H.} \bibnamefont{McAdams}} \bibnamefont{and}
  \bibinfo{author}{\bibfnamefont{A.}~\bibnamefont{Arkin}},
  \bibinfo{journal}{Trends Genet} \textbf{\bibinfo{volume}{15}},
  \bibinfo{pages}{65} (\bibinfo{year}{1999}).

\bibitem[{\citenamefont{Shen-Orr et~al.}(2002)\citenamefont{Shen-Orr, Milo,
  Mangan, and Alon}}]{SMMA2002}
\bibinfo{author}{\bibfnamefont{S.~S.} \bibnamefont{Shen-Orr}},
  \bibinfo{author}{\bibfnamefont{R.}~\bibnamefont{Milo}},
  \bibinfo{author}{\bibfnamefont{S.}~\bibnamefont{Mangan}}, \bibnamefont{and}
  \bibinfo{author}{\bibfnamefont{U.}~\bibnamefont{Alon}},
  \bibinfo{journal}{Nature genetics} \textbf{\bibinfo{volume}{31}},
  \bibinfo{pages}{64} (\bibinfo{year}{2002}).

\bibitem[{\citenamefont{Aldana and Cluzel}(2003)}]{Aldana:2003}
\bibinfo{author}{\bibfnamefont{M.}~\bibnamefont{Aldana}} \bibnamefont{and}
  \bibinfo{author}{\bibfnamefont{P.}~\bibnamefont{Cluzel}},
  \bibinfo{journal}{Proc.~Natl.~Acad.~Sci.~USA} \textbf{\bibinfo{volume}{100}},
  \bibinfo{pages}{8710} (\bibinfo{year}{2003}).

\bibitem[{\citenamefont{Klemm and Bornholdt}(2005{\natexlab{a}})}]{KB2004}
\bibinfo{author}{\bibfnamefont{K.}~\bibnamefont{Klemm}} \bibnamefont{and}
  \bibinfo{author}{\bibfnamefont{S.}~\bibnamefont{Bornholdt}},
  \bibinfo{journal}{Proc.~Natl.~Acad.~Sci.~USA} \textbf{\bibinfo{volume}{102}},
  \bibinfo{pages}{18414} (\bibinfo{year}{2005}{\natexlab{a}}).

\bibitem[{\citenamefont{Kauffman and Smith}(1986)}]{KauffmanSmith86}
\bibinfo{author}{\bibfnamefont{S.~A.} \bibnamefont{Kauffman}} \bibnamefont{and}
  \bibinfo{author}{\bibfnamefont{R.~G.} \bibnamefont{Smith}},
  \bibinfo{journal}{Physica D} \textbf{\bibinfo{volume}{22}},
  \bibinfo{pages}{68} (\bibinfo{year}{1986}).

\bibitem[{\citenamefont{Wagner}(1996)}]{Wagner:1997}
\bibinfo{author}{\bibfnamefont{A.}~\bibnamefont{Wagner}},
  \bibinfo{journal}{Evolution} \textbf{\bibinfo{volume}{50}},
  \bibinfo{pages}{1008} (\bibinfo{year}{1996}).

\bibitem[{\citenamefont{Bornholdt and Sneppen}(1998)}]{BS1998}
\bibinfo{author}{\bibfnamefont{S.}~\bibnamefont{Bornholdt}} \bibnamefont{and}
  \bibinfo{author}{\bibfnamefont{K.}~\bibnamefont{Sneppen}},
  \bibinfo{journal}{Phys. Rev. Lett.} \textbf{\bibinfo{volume}{81}},
  \bibinfo{pages}{236} (\bibinfo{year}{1998}).

\bibitem[{\citenamefont{Szejka and Drossel}(2007)}]{szejka-2007}
\bibinfo{author}{\bibfnamefont{A.}~\bibnamefont{Szejka}} \bibnamefont{and}
  \bibinfo{author}{\bibfnamefont{B.}~\bibnamefont{Drossel}}  
  \bibinfo{journal}{Eur. Phys. J. B} \textbf{\bibinfo{volume}{56}},
  \bibinfo{pages}{373-380} (\bibinfo{year}{2007}). 

\bibitem[{\citenamefont{Ciliberti et~al.}(2007)\citenamefont{Ciliberti, Martin,
  and Wagner}}]{Ciliberti:2007lr}
\bibinfo{author}{\bibfnamefont{S.}~\bibnamefont{Ciliberti}},
  \bibinfo{author}{\bibfnamefont{O.~C.} \bibnamefont{Martin}},
  \bibnamefont{and} \bibinfo{author}{\bibfnamefont{A.}~\bibnamefont{Wagner}},
  \bibinfo{journal}{PLoS Comp. Biol.} \textbf{\bibinfo{volume}{3}}
  (\bibinfo{year}{2007}).

\bibitem[{\citenamefont{Greil and Drossel}(2005)}]{greil:048701}
\bibinfo{author}{\bibfnamefont{F.}~\bibnamefont{Greil}} \bibnamefont{and}
  \bibinfo{author}{\bibfnamefont{B.}~\bibnamefont{Drossel}},
  \bibinfo{journal}{Phys. Rev. Lett.} \textbf{\bibinfo{volume}{95}},
  \bibinfo{eid}{048701} (\bibinfo{year}{2005}).

\bibitem[{\citenamefont{Klemm and Bornholdt}(2005{\natexlab{b}})}]{KB2004-2}
\bibinfo{author}{\bibfnamefont{K.}~\bibnamefont{Klemm}} \bibnamefont{and}
  \bibinfo{author}{\bibfnamefont{S.}~\bibnamefont{Bornholdt}},
  \bibinfo{journal}{Phys. Rev. E} \textbf{\bibinfo{volume}{72}},
  \bibinfo{pages}{055101} (\bibinfo{year}{2005}{\natexlab{b}}).

\bibitem[{\citenamefont{Braunewell and Bornholdt}(2007)}]{Braunewell:2007lr}
\bibinfo{author}{\bibfnamefont{S.}~\bibnamefont{Braunewell}} \bibnamefont{and}
  \bibinfo{author}{\bibfnamefont{S.}~\bibnamefont{Bornholdt}},
  \bibinfo{journal}{J.~Theor.~Biol.} \textbf{\bibinfo{volume}{245}},
  \bibinfo{pages}{638} (\bibinfo{year}{2007}).

\bibitem[{\citenamefont{Hirata et~al.}(2002)\citenamefont{Hirata, Yoshiura,
  Ohtsuka, Bessho, Harada, Yoshikawa, and Kageyama}}]{Hi2002}
\bibinfo{author}{\bibfnamefont{H.}~\bibnamefont{Hirata}},
  \bibinfo{author}{\bibfnamefont{S.}~\bibnamefont{Yoshiura}},
  \bibinfo{author}{\bibfnamefont{T.}~\bibnamefont{Ohtsuka}},
  \bibinfo{author}{\bibfnamefont{Y.}~\bibnamefont{Bessho}},
  \bibinfo{author}{\bibfnamefont{T.}~\bibnamefont{Harada}},
  \bibinfo{author}{\bibfnamefont{K.}~\bibnamefont{Yoshikawa}},
  \bibnamefont{and} \bibinfo{author}{\bibfnamefont{R.}~\bibnamefont{Kageyama}},
  \bibinfo{journal}{Science} \textbf{\bibinfo{volume}{298}},
  \bibinfo{pages}{840} (\bibinfo{year}{2002}).

\bibitem[{\citenamefont{Wuensche}(1994)}]{Wuensche:1994}
\bibinfo{author}{\bibfnamefont{A.}~\bibnamefont{Wuensche}},
  \emph{\bibinfo{title}{Artificial Life III}}
  (\bibinfo{publisher}{Addison-Wesley, Reading, MA}, \bibinfo{year}{1994}),
  chap. \bibinfo{chapter}{The Ghost in the Machine: Basins of Attraction of
  Random Boolean Networks}.

\end{thebibliography}
\end{document}